\documentclass[useAMS,usenatbib]{mn2e}
\usepackage{amsfonts}
\usepackage{epsfig}
\usepackage{graphicx}

\title[A Spectroscopic Study of the Blue Stragglers in M67]{A Spectroscopic
  Study of the Blue Stragglers in M67}

\author[G. Q. Liu et al.]{G. Q. Liu,$^{1,2}$\thanks{E-mail: lgq@bao.ac.cn}
L. Deng,$^{1}$\thanks{E-mail: licai@bao.ac.cn}
M. Ch\'avez,$^{3}$\thanks{E-mail: mchavez@inaoep.mx}
E. Bertone,$^{3}$\thanks{E-mail: ebertone@inaoep.mx}
A. Herrero Davo,$^{4}$\thanks{E-mail: ahd@iac.es}
\newauthor
and M. D. Mata-Ch\'avez$^{5}$\\
$^1$NAOC -- National Astronomical Observatories, Chinese Academy of
Sciences, Beijing 100012, P. R. China\\
$^2$GUCAS -- Graduate University of Chinese Academy of Sciences, Beijing
100049, P. R. China\\
$^3$INAOE -- Instituto Nacional de Astrof\'\i sica, \'Optica y
     Electr\'onica, Luis Enrique Erro 1, 72840 Tonantzintla, Puebla, Mexico\\
$^4$IAC -- Instituto de Astrof\'\i ca de Canarias, E38205 La Laguna, Tenerife, Spain\\
$^5$Departamento de Fisica, CUCEI, Universidad de Guadalajara, Blvd. Marcelino
Garcia Barragan 1412, Guadalajara, Jalisco, Mexico}

\date{Received date; accepted date}
\pubyear{2008}

\begin{document}
\maketitle

\begin{abstract}
Based on spectrophotometric observations from the
Guillermo Haro Observatory (Cananea, Mexico), a study of the spectral
properties of the complete sample of 24 blue straggler stars (BSs) in
the old Galactic open cluster M67 (NGC2682) is presented. All
spectra, calibrated using spectral standards, were re-calibrated by
means of photometric magnitudes in the
Beijing-Arizona-Taipei-Connecticut system, which includes fluxes in 11
bands covering $\sim 3500-10000$~\AA. The set of parameters was
obtained using two complementary approaches that rely on a comparison
of the spectra with (a) an empirical sample of stars with
well-established spectral types and (b) a theoretical grid of optical
spectra computed at both low and high resolution. The overall results
indicate that the BSs in M67 span a wide range in $T_{\rm eff}$ ($\sim
5600-12600$~K) and surface gravities that are fully compatible with
those expected for main sequence objects ($\log{g}=3.5-5.0$~dex).
\end{abstract}


\section{Introduction}

Blue straggler stars (BSs) were first discovered in the globular
cluster M3 by Sandage (1953). These peculiar stars were named `blue
stragglers' because of their observational properties in star cluster
colour--magnitude diagrams (CMDs). Usually, BSs appear as a bluer and
brighter extension of a cluster's main sequence (MS). As members of
the same star cluster and having been born at the same time, the
behaviour of BSs is paradoxical because there should be no main
sequence stars above the turn-off according to the standard
theoretical picture of stellar evolution in such a coeval and
initially chemically homogeneous system.

Decades have passed since their discovery, in which they have been the
subject of many studies. These peculiar stars have been found to be
common constituents of virtually all evolved systems (and also in
young systems, but a `normally populated main sequence' would hide any
BSs), including dwarf galaxies (Stryker 1993). Based on observational
and theoretical studies, it is generally believed that the BSs in
high-density regions of stellar systems could be the remnants of
stellar collisions and those in sparse environments might result from
the coalescence of interacting binaries or mass transfer through
Roche-lobe overflow in primordial binary systems (Ahumada 1999; Bacon
et al. 1996; Ferraro et al. 1997; Gilliland \& Brown 1992; Leonard 1989;
Livio 1993; Ouellette \& Pritchet 1998; Piotto et al. 1999; Stryker
1993; Tian et al. 2006). In addition to their still elusive origin,
the study of BSs is important because in a stellar population they are
among the most massive and luminous stars, whose contribution to the
integrated light cannot be predicted by the standard theory of stellar
evolution (Bressan et al. 1993). In fact, it has been demonstrated
that they greatly affect the spectral energy distribution (SED) of the
entire population (Deng et al. 1999; Manteiga et al. 1989),
particularly at ultraviolet and blue wavelengths (Xin et
al. 2007, 2008).

In spite of the numerous studies published since their discovery, it
is still not clear which of the conceivable explanations for the BS
phenomenon is the preferred (or dominant) mechanism of formation.
Similarly, it has not yet been established whether the spectral
properties of BSs are the same as those of regular main sequence stars
of the same mass, as would be expected according to their loci in the
CMDs, although this is in contrast to the potential chemical
enrichment in the atmospheres presumably provoked by the different
detailed formation processes.

\begin{table}
\caption{The blue straggler population of M67. `n' is the number
of spectra collected for each object.
\label{basic}}
\begin{tabular}{ccccl}
\hline\hline
Name & R.A.(2000) & Dec.(2000)
& ExpTime(s) & n\\
\hline
BS005 &8:51:11.78 &11:45:22.24 &2400 &4\\
BS018 &8:52:10.75 &11:44:06.07 &1200 &2\\
BS025 &8:51:27.04 &11:51:52.22 &1200 &3\\
BS029 &8:51:48.65 &11:49:15.36 &2400 &4\\
BS034 &8:51:34.31 &11:51:10.23 &2400 &4\\
BS038 &8:51:32.61 &11:48:52.02 &1200 &2 \\
BS040 &8:51:26.45 &11:43:50.75 &1200 &2\\
BS043 &8:51:14.37 &11:45:00.70 &2400 &4\\
BS046 &8:51:20.82 &11:53:25.65 &2400 &4\\
BS047 &8:51:03.52 &11:45:02.68 &1200 &2\\
BS065 &8:51:21.77 &11:52:38.00 &2400 &4\\
BS093 &8:51:32.57 &11:50:40.42 &1200 &2 \\
BS111 &8:51:19.92 &11:47:00.50 &1500 &2\\
BS115 &8:51:37.72 &11:37:03.54 &1200 &2 \\
BS116 &8:50:55.70 &11:52:14.50 &2400 &4\\
BS126 &8:49:21.49 &12:04:23.00 &1200 &2\\
BS131 &8:51:28.40 &12:07:38.30 &1200 &2 \\
BS139 &8:51:39.24 &11:50:03.66 &1500 &2\\
BS143 &8:51:21.25 &11:45:52.63 &1200 &2 \\
BS182 &8:51:15.47 &11:47:31.74 &1800 &2\\
BS184 &8:50:47.69 &11:44:51.33 &3300 &4\\
BS185 &8:51:28.17 &11:49:27.06 &3000 &4\\
BS206 &8:48:59.84 &11:44:51.66 &600  &1\\
BS216 &8:51:20.59 &11:46:16.36 &1500 &2\\
\hline
\end{tabular}
\end{table}

Nevertheless, BSs have historically been regarded as
core-hydrogen-burning stars (Benz \& Hills 1987, 1992). For this
reason, it is usually assumed that the spectral properties of BSs are
compatible with those of main sequence stars at the same loci in the
CMDs. We have adopted this assumption throughout a recent series of
papers discussing the integrated SEDs (ISEDs) of star clusters at low
spectral resolution (Deng et al. 1999; Xin \& Deng 2005; Xin et al. 2007;
Xin et al. 2008). However, whether BSs can actually be
represented by main sequence objects has not yet been fully
investigated, perhaps with the exception of relatively few early
papers (Strom et al. 1971). The present paper is
therefore aimed at validating this assumption observationally.
Determinations are also obtained of the two fundamental parameters,
i.e., the effective temperature and the surface gravity, of the full
sample of BSs in M67, based on a homogeneous collection of spectra.

With the purpose of properly assessing their nature, we started a
long-term project aimed at determining the atmospheric parameters of
BSs in stellar systems. We will first determine the effective
temperatures and surface gravities of the objects, through photometric
and intermediate-resolution spectroscopic observations. In a second
step we will investigate the chemical details for (corroboration of) a
possible binary nature and to establish the existence (or absence) of
the chemical patterns associated with a mass-transfer process.

In this paper we present the initial steps of this project by
investigating the full sample of BSs in the well-studied old open
cluster M67 (NGC2682). M67 contains a rich system of 24 BSs (Deng et
al. 1999), a sample sufficiently large for statistical purposes. The
present paper is organised as follows. In Section~2 we describe the
observations. In Section~3 we give the details of the flux-fitting
method and provide the final sets of parameters. Fine-tuning
of the gravity determination is described in Section 4. In Section 5,
a comparison with previous work is presented. Finally, a summary and
the conclusions of this study are presented in Section 6.

\section{Observations and Reduction}

The observations were carried out during a three-night run in
February 2005 using the 2.12 m telescope of the Guillermo Haro
Observatory (OAGH) at Cananea, Mexico. The spectra were collected using the
Boller \& Chivens spectrograph with a 150 $\ell$/mm grating blazed at
5000 {\AA} and a Tektronix 1024$\times$1024 CCD detector. The
instrumental set-up yielded a scale of 3.2 {\AA} per pixel with a
wavelength coverage roughly from 3600 to 6900 {\AA} at a nominal 5.7
{\AA} full width at half-maximum (FWHM), with a slit width of 150
$\mu$m. A total of 66 object frames were observed, which included at
least two frames per object, with the exception of BS206 for which
we were able to observe only once.

\begin{figure*}
\includegraphics[width=21cm]{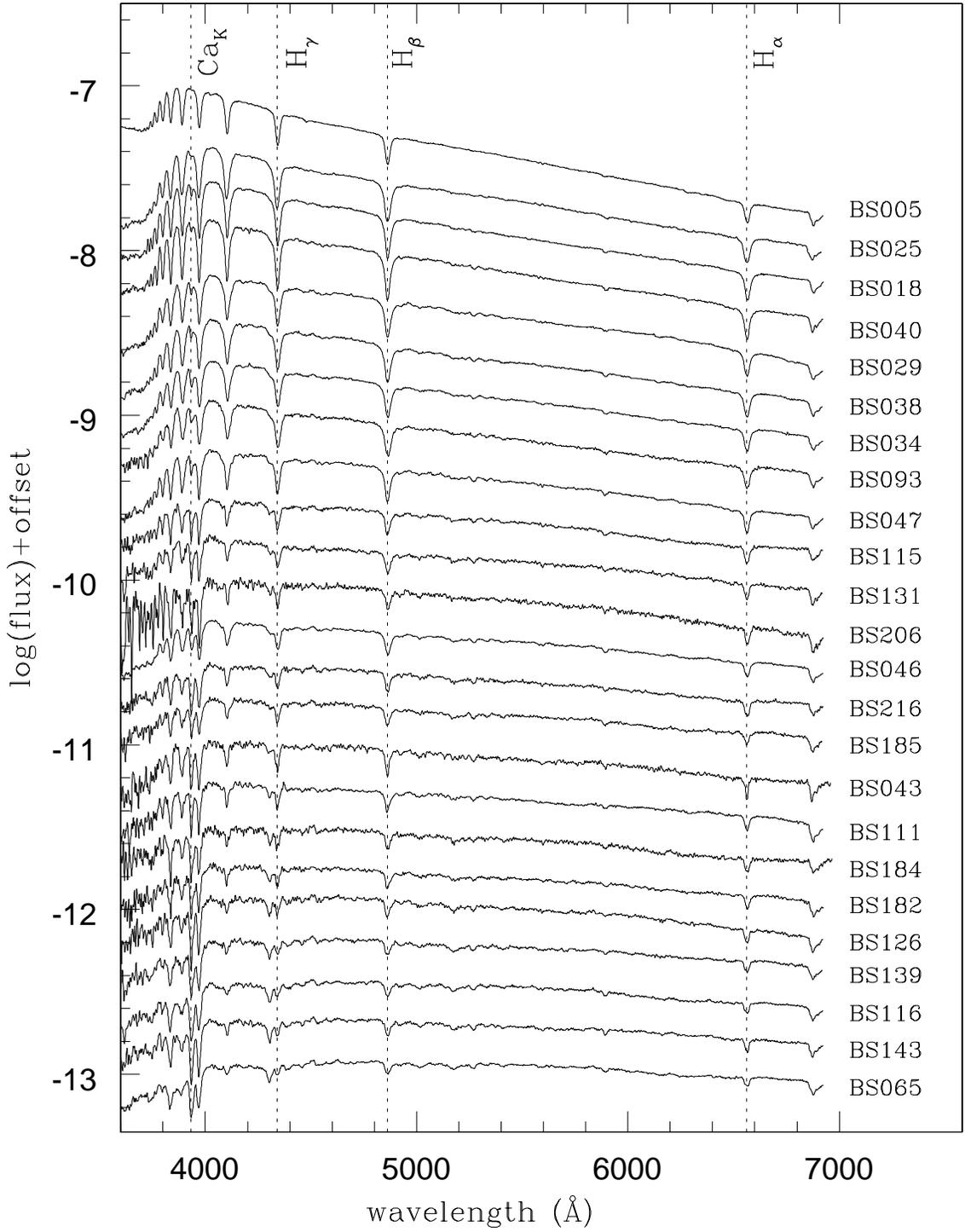}
\caption{Observed spectral energy distributions of the 24 BSs in our
sample. The spectra are roughly ordered in a temperature sequence,
decreasing from top to bottom. Vertical dotted lines indicate the loci
of four major spectral features, H${\alpha}$ 6562~{\AA}, H${\beta}$
4860~{\AA}, H${\gamma}$ 4340~{\AA} and Ca K 3933~{\AA}.}\label{all}
\end{figure*}

\begin{table}
\caption {Artificial colours and magnitudes of our sample BSs.\label{bvv}}
\begin{center}
\begin{tabular}{ccr}
\hline\hline
Name & $V$ & $B-V$ \\
\hline
BS005 &10.02 &$-$0.064\\
BS018 &10.68 &0.083\\
BS025 &10.95 &0.107\\
BS029 &10.93 &0.200\\
BS034 &10.95 &0.232\\
BS038 &11.10 &0.190\\
BS040 &11.29 &0.118\\
BS043 &10.94 &0.437\\
BS046 &11.22 &0.405\\
BS047 &11.34 &0.301\\
BS065 &11.32 &0.613\\
BS093 &12.27 &0.247\\
BS111 &12.16 &0.447\\
BS115 &12.33 &0.375\\
BS116 &11.99 &0.576\\
BS126 &12.22 &0.492\\
BS131 &12.60 &0.385\\
BS139 &12.28 &0.538\\
BS143 &12.30 &0.556\\
BS182 &12.71 &0.474\\
BS184 &12.72 &0.482\\
BS185 &12.82 &0.423\\
BS206 &12.92 &0.396\\
BS216 &12.92 &0.430\\
\hline
\end{tabular}
\end{center}
\end{table}

\begin{figure}
\begin{center}
\includegraphics[width=8cm]{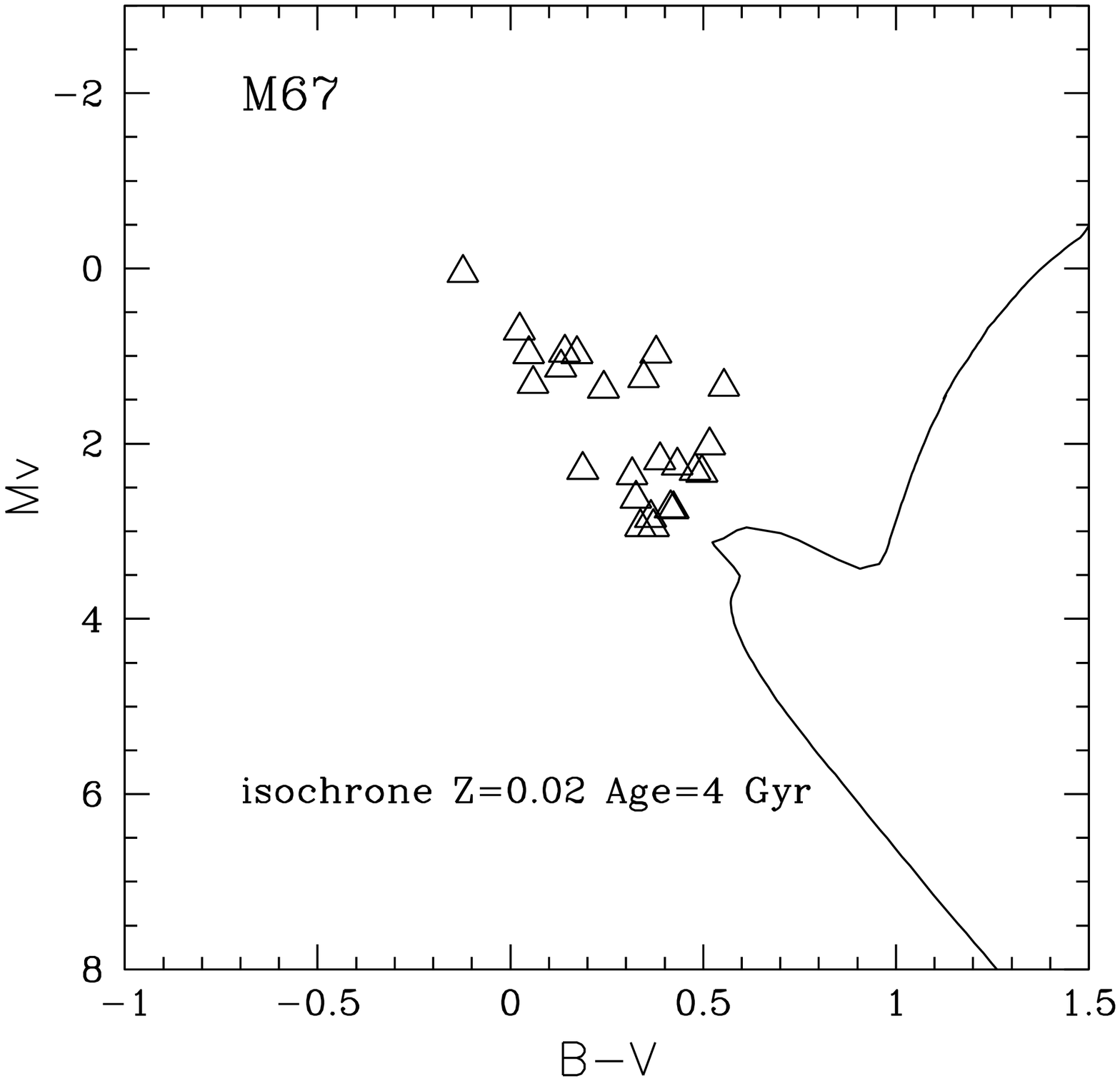}
\end{center}
\caption{Artificial colour--magnitude diagram for the BSs in
M67. The full sample of 24 BSs are marked with open triangles. The
solid line is the 4 Gyr isochrone of solar metallicity from Bertelli
et al. (1994).}\label{cmd}
\end{figure}

For the data reduction we followed standard procedures using the IRAF
package. Bias and flat-field corrections were secured by collecting a
set of ten bias frames as well as dome-projected halogen-lamp images
at the beginning and end of each night. Each stellar image was
accompanied by a Helium-Argon-lamp image that allowed wavelength
calibration and the determination of the nominal resolution along the
dispersion axis. The relative flux calibration was done using
observations of three spectrophotometric standard stars, BD75325,
Feige67, and Feige34.

The 24 BSs in our sample are all members of M67 with nearly 100 per cent
membership probabilities, as determined from both proper-motion and
radial-velocity observations (Girard et al. 1989; Sanders 1977). The
catalogue is included in Table~\ref{basic} where we give in columns (1)
to (5) the BS identification numbers from Fan et al. (1996), the
equatorial coordinates, the integrated exposure times (in seconds),
and the number of spectra collected for each object. The resulting
relative-flux-calibrated spectra are shown in Fig.~\ref{all}.

Qualitatively, the spectra in Fig.~\ref{all} are roughly ordered as a
temperature sequence, based on visual inspection of the slope of the
SED. It is interesting to note that the sequence of Balmer features,
distinguishable down to H11 ($\lambda=$3771\AA), from bottom to
top, exhibits an increase up to BS025, indicating that BS005 should
have a temperature compatible with that expected for a late-B
star. Similarly, the Ca K line at 3933~\AA\ is nearly absent in the
two hottest objects and steadily increases in strength, overcoming the
intensity of the blend with H$\epsilon$ and the Ca H line at 3968~\AA\
at the position of BS184. Another interesting feature easily
observable in the spectra is the CH G-band at 4300~\AA. This
feature is strongest for the two objects at the bottom. Again, from
bottom to top, this feature disappears at the position of
BS093. Therefore, this star should correspond roughly to spectral type
A5 (about $T_{\rm eff}=8000$~K).

\begin{figure}
\begin{center}
\includegraphics[width=8cm]{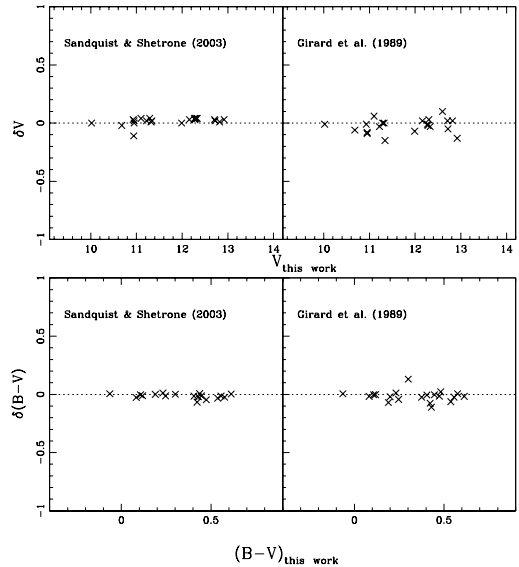}
\end{center}
\caption{A comparison of visual magnitudes (top panels) and $B-V$
colour indices (bottom panels) of our programme stars between the
present paper and those of Sandquist \& Shetrone (2003) and Girard et
al. (1989).}\label{comp}
\end{figure}

We apply an absolute-flux calibration using intermediate-band
photometric data (resembling spectrophotometric observations)
collected at an earlier time (Deng et al. 1999). In brief, the 24 BSs
were observed photometrically using the
Beijing-Arizona-Taipei-Connecticut (BATC) intermediate-band
filters. These observations included 11 of the 15 filters in this
system, covering a range from 3500 to 10000 {\AA}. The shortest
wavelength covered by our dataset was obtained using a filter centered
on 3890 {\AA}. By convolving the observed BS spectra with the six
intermediate-band ($b, d, f, g, h$, and $i$; central wavelengths at
3890 {\AA}, 4550 {\AA}, 5270 {\AA}, 5795 {\AA}, 6075 {\AA}, and 6660
{\AA}, respectively) transmission curves, six new magnitudes for each
object were derived. These magnitudes were compared with those
obtained with the BATC photometric observations and permitted the
derivation of the scaling factors to transform the spectroscopy-based
magnitudes to the absolute BATC system. The accurate
intermediate-band photometry in the BATC system secures
(re-calibrates) the overall shape of the observed spectra for all
programme stars.

\begin{figure*}
\begin{center}
\includegraphics[width=15cm]{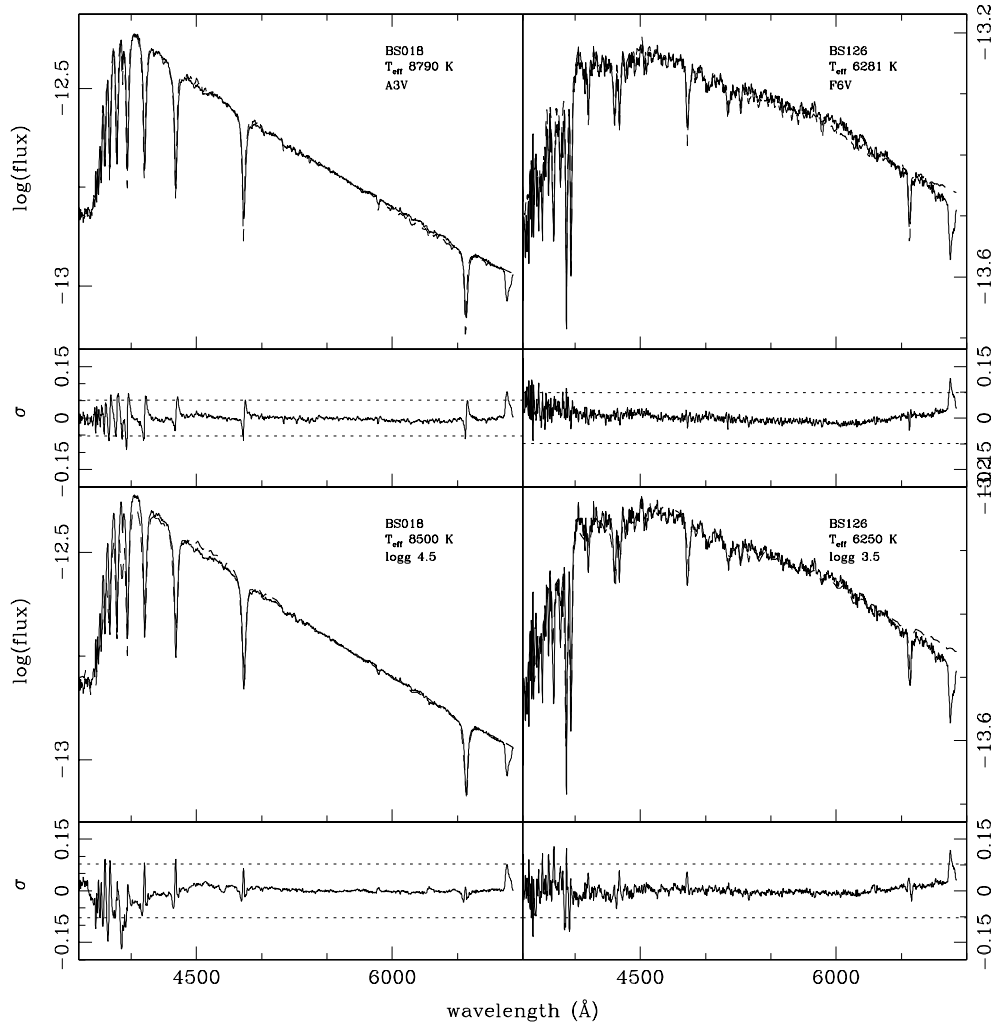}
\end{center}
\caption{Best fits for two representative BSs, BS018 and BS126.  The
solid and dashed lines show, respectively, the observed spectra and
the best-fit Pickles (top panels) and Kurucz low-resolution spectra
(bottom panels). The smaller panels below each spectrum correspond to
the residuals, as explained in the text, and the dotted lines indicate
the 3$\sigma$ boundaries.}\label{fit}
\end{figure*}

To assess the quality of the overall SED shapes and to check the
precision of the calibration, the spectrophotometrically calibrated
spectra were used to construct broad-band photometry, which can then
be compared with standard broad-band observations. Two sets of $V,
(B-V)$ observations from Sandquist \& Shetrone (2003) -- who
provided photometric data of BSs in M67 for a study of variability
in the light curves -- and Girard et al. (1989) -- who studied the
relative proper motions and the stellar velocity dispersion of M67
-- are compared with the artificial magnitudes and colours (see
Table~\ref{bvv}) of the 24 BSs, derived by convolving the $B$ and
$V$ filter-response functions with the calibrated spectra. Adopting
a distance modulus of DM=9.97 nag and a colour excess $E(B-V)$
of 0.059 mag for M67 (taken from WEBDA\footnote
{http://www.univie.ac.at/webda/}), the observations (both the
location with respect to the cluster's main sequence turn-off and
the magnitudes and colours of the programme stars) can very well be
reproduced using the artificial CMD photometry, as shown in
Fig.~\ref{cmd}. The artificial photometry was also compared with
direct broad-band photometry (magnitudes and colours). A perfect
match was found, as shown in Fig.~\ref{comp}, in which the
residuals in $V$ magnitude, $\delta$ V=V-V$_{\rm this~ work}$, and
$(B-V)$ colour index $\delta(B-V)=(B-V)-(B-V)_{\rm this~ work}$ are
displayed. No systematic differences were found, whereas the random
difference between the data derived from our spectra and from direct
photometry is compatible with observational errors of a few per cent
of a magnitude. Figure~\ref{comp} independently shows that our
spectral observations and calibration are accurate to a satisfactory
degree.

\section{Spectral Fitting and Analysis}

In order to study the spectral properties of BSs and to
determine the effective temperature and surface gravity of our sample
stars, we applied simple flux-fitting methods, using three different
libraries of reference stellar spectra, both observed and
synthetic. In all cases we assume a solar chemical composition for M67
stars, which is in agreement with observational determinations
(Bressan \& Tautvai\~{s}iene 1996; Hobbs \& Thorburn 1991).

\begin{figure}
\begin{center}
\includegraphics[width=5cm]{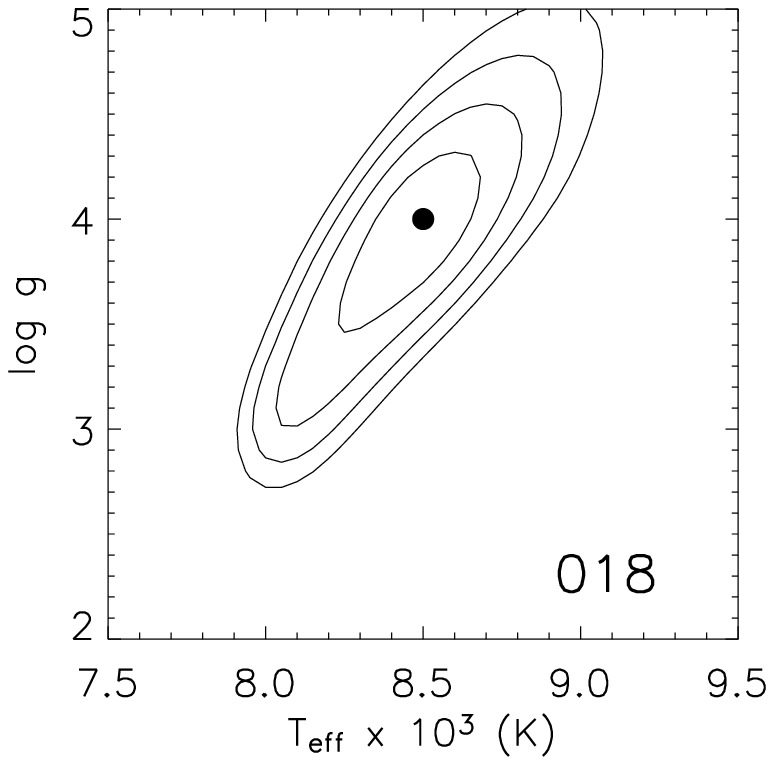}
\includegraphics[width=5cm]{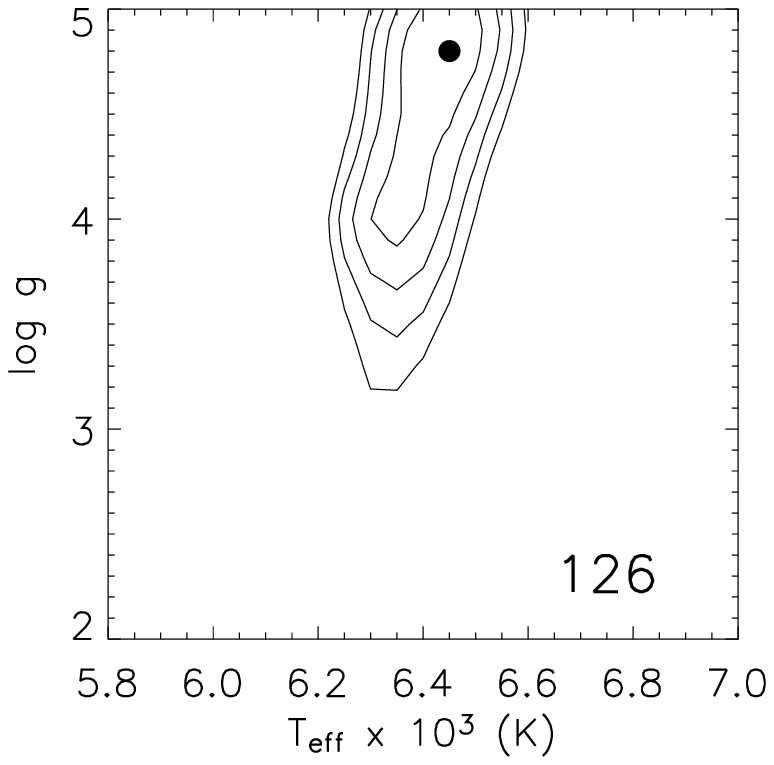}
\end{center}
\caption{Contour plots for BS018 ({\it top panel}) and BS126 ({\it
bottom panel}). The solid dot corresponds to the best parameter
estimate; the contour levels indicate the 1, 2, 3, and 4$\sigma$
uncertainties.\label{contour}}
\end{figure}

\begin{enumerate}
\item Each calibrated spectrum was compared with every entry in the
spectral atlas of Pickles (1998). The comparison was carried out after
normalising our spectra and those in the reference atlas to the flux
at $\lambda=$5556 {\AA}. The algorithm we have implemented finds the
spectrum (and its associated parameters) in the atlas that produces
the minimum standard deviation, $\sigma$, of the residual flux,
$\Delta F = F_{\rm BS}-F_{\rm Pickles}$, computed for each $\lambda$.

Because of the marked decrease in sensitivity of the CCD at the
shortest wavelengths, the spectral regime considered for the flux
fitting excludes the region at $\lambda <$ 3850
{\AA}. Figure~\ref{fit} displays the best fit and residuals for BS018
and BS126 ({\it top panels}). The solid and dashed lines are,
respectively, the BSs' calibrated spectra and the best-fit flux from
the Pickles library. The labels at the top indicate the star ID, its
temperature and spectral-type designation.

\item A similar procedure was followed by using the spectral grids of
Lejeune et al. (1997, 1998) which are, for the segment of the
parameter space under consideration, mostly based on Kurucz (1993)
low-resolution theoretical fluxes. In this case, both sets of spectra
(BSs and model fluxes) were normalised to the flux at $\lambda =$ 5390
{\AA}. A set of best-fit parameters is found by directly comparing the
observed spectra with each of the model fluxes. It is important to
note that in this way, as well as in the previous point, the best fit
always corresponds to a grid point. In Fig.~\ref{fit} we show the best
fit for the stars BS018 and BS126 ({\it bottom panels}). The solid and
dashed lines are, respectively, the BSs' calibrated spectra and the
best-fit theoretical flux. The label on the right gives the parameters
of the best-fit model atmosphere.

\item In this case we made use of the {\sc bluered} library
(Bertone et al. 2003, 2008). {\sc bluered} is a high-resolution
($R$=500,000) grid of over 800 synthetic stellar spectra, covering
SEDs in the optical range ($\lambda$ = 3500$-$7000 {\AA}). The library
is based on the ATLAS9 model atmospheres and has been computed with
the SYNTHE code developed by Kurucz (1993). The grid spans a large
volume in the fundamental parameter space, accounting for virtually any
stellar type from O to M stars and from dwarfs to supergiants. An
important aspect of this grid, although of marginal relevance for the
parameters associated with our programme stars, is that its
calculation includes the effect of diatomic molecules, in particular
TiO. A best-fit spectrum was found in the two-dimensional space
covering ($T_{\rm eff}$, $\log{g}$), after minimising the statistical
variance in the relative-flux domain as a measure of the similarity
between target spectrum and theoretical SEDs across {\sc bluered}
(Bertone et al. 2004). As in the comparisons above, we have assumed a
solar chemical composition for M67. It is worth noting that the grid
of theoretical spectra has been properly modified to simulate the
instrumental set-up.
The results are shown in columns (6) and (7) of Table~\ref{result},
whereas contour plots for BS018 and BS126 are shown in
Fig.~\ref{contour}.
\end{enumerate}

The results from the three methods are listed in Table~\ref{result},
where columns (1) to (8) include, respectively, the object ID, the
parameter pairs (T$_{\rm eff}$, log $g$ or spectral type) and
the identification numbers following Sanders (1977), for ease of
cross identification.

The agreement among the effective temperature estimates provided by
the three methods is on the order of 2--6 per cent and, in general,
the best-fit T$_{\rm eff}$ values based on the {\sc bluered}
library are the highest, apart from the case of the hottest star,
BS005, where the best fit is about 1550 K lower. The discrepancy in
this case arises from the associated low log $g$ value, which is about
2 dex lower than the corresponding result from the Lejeune library,
since the flux-fitting method, applied to intermediate-resolution
spectra, is affected by a T$_{\rm eff}$$-$log $g$ degeneracy, where a
lower surface gravity implies a cooler temperature (Buzzoni et
al. 2001).

A larger discrepancy affects the derived surface gravities of our
sample stars, for which the highest values are most often provided
by the {\sc bluered} library. These latter results are, however,
affected by an average 1$\sigma$ error of $\la 1$ dex (see
%
%
Fig.~\ref{contour}.) The systematically higher parameter values
of the {\sc bluered} spectra can be understood as in Bertone et
al. (2008), who show that the T$_{\rm eff}$ and log $g$ values that
are obtained from comparing the Sun with the {\sc bluered} spectra
(at very high spectral resolution) are a few per cent higher than
those commonly accepted because the physical parameters of the
absorption lines included in the spectral synthesis generate deeper
features -- which are counterbalanced by raising both the effective
temperature and the surface gravity. In general, the current
determinations of gravity for the programme stars are limited by the
low resolution.

%
%

\begin{table*}
\begin{center}
\caption{Best-fit parameters.\label{result}}
\begin{tabular}{lrlrrrrr}
\hline\hline $Name^a$ &
\multicolumn{2}{c}{BATC-Pickles}  &
\multicolumn{2}{c}{BATC-Lejeune}  &
\multicolumn{2}{c}{BATC-\sc{bluered}} & $S^b$  \\
&$T_{\rm eff}$ &Sp.Type &$T_{\rm eff}$ & $\log{g}$ & $T_{\rm eff}$ &
$\log{g}$ & \\ &(K)& &(K)&(dex)&(K) & (dex) & \\
\hline
BS005     &12589&B6IV&12625&5.00&11050&3.1&977 \\
BS018     &8790&A3V  &8500 &4.50&8500&4.0&1434\\
BS025     &8492&A5V  &8500 &4.25 &8950&5.0&1066 \\
BS029$^*$ &8054&A7V  &7813 &4.38 &8100&5.0&1267\\
BS034$^*$ &8054&A7III&7688 &4.38 &7900&5.0&1284\\
BS038     &8054&A7III&7750 &4.00 &8050&5.0&1263 \\
BS040     &8790&A3V  &8625 &4.75 &8450&4.2&968 \\
BS043$^*$ &6469&F5V  &6500 &3.50 &6700&4.7&975 \\
BS046$^*$ &6776&F2V  &6625 &3.88 &6850&4.7&1082 \\
BS047$^*$ &7586&F0III&7250 &4.00 &7500&4.8&752\\
BS065     &5636&G2V  &5750 &3.50 &6000&4.2&1072  \\
BS093     &7586&F0III&7625 &4.25 &7800&5.0&1280\\
BS111$^*$ &6281&F6V  &6500 &3.50 &6600&4.6&997 \\
BS115$^*$ &6776&F2V  &7000 &4.25 &7050&4.7&1195\\
BS116     &6039&F8V  &5938 &3.75 &6150&4.5&792\\
BS126     &6281&F6V  &6250 &3.50 &6450&4.8&277 \\
BS131     &6776&F2V  &6875 &4.75 &6950&4.7&1273  \\
BS139     &6039&F8V  &6000 &3.50 &6200&4.6&984 \\
BS143     &6039&F8V  &6000 &3.50 &6100&4.1&1005 \\
BS182     &6281&F6V  &6250 &4.00 &6500&4.7&751 \\
BS184$^*$ &6281&F6V  &6375 &4.25 &6500&4.6&1036  \\
BS185     &6531&F5V  &6438 &4.00 &6700&4.7&145 \\
BS206     &6776&F2V  &6750 &4.50 &6900&4.7&2204\\
BS216     &6531&F5V  &6500 &4.50 &6700&4.7&2226 \\
\hline\hline
\end{tabular}
\end{center}
{\sc NOTE:} a, Stellar Identification from Fan et al. (1996); b,
Sanders number (Sanders 1977); *, Binary population.
\end{table*}

\section{Fine-tuning of the gravity determination}

Complementary to the analysis presented in the previous section, we
obtained observations at OAGH of the BS sample with an
alternative set-up that allows, in principle, the separation of
potentially fiducial gravity indicators. In particular, we will make
use of the indicators defined by Rose (1984, 1994), which consist of
line ratios of several pairs of features and the corresponding
index-index diagnostic diagrams, and the hydrogen-absorption indices
defined by Worthey et al. (1994) as part of the Lick system. These two
approaches are necessary in view of the large effective temperature
interval covered by the BSs.

New observations were carried out on February 24--27, 2008, using the
Boller \& Chivens spectrograph and the Versarray 1300$\times$1300 CCD
detector optimised for the blue spectral interval. We used the
600~$\ell$/mm grating and a slit width of 200~$\mu$m, which yielded a
nominal dispersion of 0.7 \AA/pixel and a resolution of 2.6~\AA\
FWHM. The grating was positioned to obtain spectra in the interval
3800--4700~\AA\, where all of the gravity indicators cited above are
defined.

The sample consisted of two stellar sets. The first corresponds to the
full sample of BSs, whereas the other contains nearly 50 objects that
served as gravity templates. The latter set was selected from the
catalogues of Cayrel de Strobel et al. (1997, 2001). Data reduction up
to the standard flux calibration was performed using the conventional
procedures of IRAF and utilising a set of standard stars observed each
night. We considered it very important to secure calibrated fluxes to
provide reproducible results when analysing data collected with other
instruments. In Table~\ref{tab:refstars} we list the control sample,
showing in columns (1) to (5) the stellar identification, the spectral
type and the associated stellar parameters collected from Cayrel
de Strobel et al. (1997, 2001). For the stars with multiple
determinations we provide the average values.

\begin{figure}
\begin{center}
\includegraphics[width=8.5cm]{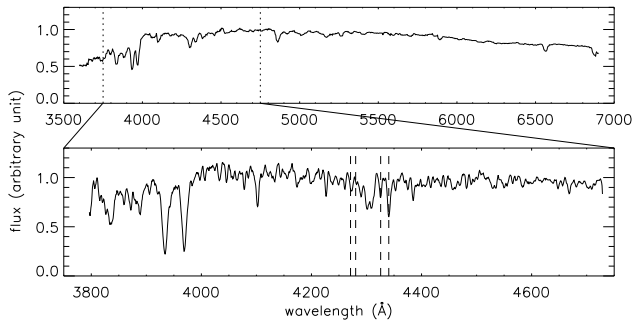}
\end{center}
\caption{Low- ({\it top panel}) and intermediate-resolution ({\it
    bottom panel}) spectra of BS065. The dotted vertical lines
    indicate the positions of the Rose (1994) features used to define
    the gravity-sensitive line-depth index used in this
    paper.\label{fig:spectrum}}
\end{figure}

As an example of the flux-calibrated spectra we display, in
Fig.~\ref{fig:spectrum}, the lower-resolution spectra of BS065 and a
{\it zoomed-in} region at higher resolution. The vertical dashed lines
indicate the position of several of the features used as gravity
indicators, as described below.

\begin{table}
\begin{center}
\caption{Stars used as gravity templates.\label{tab:refstars}}
\begin{tabular}{llrrr}
\hline\hline
Name & Sp.Type & $T_{\rm eff}$ & $\log{g}$ & [Fe/H] \\
     & & (K) & (dex) & (dex) \\
\hline
HD025621 & F6IV      &   6251 & 3.95 &    0.01 \\
HD027962 & A2IV      &   9000 & 4.00 &    0.40 \\
HD028271 & F7V       &   6160 & 3.85 & $-$0.10 \\
HD028978 & A2V       &   9164 & 3.70 &    0.14 \\
HD031295 & A0V       &   8860 & 4.12 & $-$1.08 \\
HD032537 & F0V       &   6904 & 4.00 & $-$0.30 \\
HD033256 & F2V       &   6219 & 3.94 & $-$0.31 \\
HD033608 & F5V       &   6526 & 4.09 &    0.23 \\
HD033959 & A9IV      &   7670 & 3.55 &    0.00 \\
HD034578 & A5II      &   8300 & 1.85 &    0.16 \\
HD035497 & B7III     &  13622 & 3.80 & $-$0.10 \\
HD035984 & F6III     &   6175 & 3.68 & $-$0.07 \\
HD038899 & B9IV      &  10903 & 4.00 &    0.01 \\
HD043386 & F5IV-V    &   6480 & 4.27 & $-$0.06 \\
HD061295 & F6II      &   6925 & 3.00 &    0.25 \\
HD076292 & F3III     &   6866 & 3.77 & $-$0.22 \\
HD085235 & A3IV      &  11200 & 3.55 & $-$0.40 \\
HD087822 & F4V       &   6597 & 4.10 &    0.17 \\
HD091752 & F3V       &   6352 & 3.94 & $-$0.27 \\
HD094028 & F4V       &   5960 & 4.23 & $-$1.46 \\
HD095418 & A1V       &   9953 & 4.10 &    0.47 \\
HD097633 & A2V       &   9395 & 3.57 &    0.04 \\
HD099028 & F4IV      &   6739 & 3.98 &    0.06 \\
HD099285 & F2V       &   6599 & 3.84 & $-$0.22 \\
HD100563 & F5V       &   6401 & 4.31 &    0.05 \\
HD101606 & F4V       &   6105 & 4.10 & $-$0.78 \\
HD102574 & F7V       &   6030 & 3.92 &    0.16 \\
HD110411 & A0V       &   8970 & 4.36 & $-$1.00 \\
HD117361 & F0IV      &   6789 & 3.95 & $-$0.27 \\
HD120136 & F6IV      &   6430 & 4.19 &    0.25 \\
HD126660 & F7V       &   6338 & 4.29 & $-$0.05 \\
HD128167 & F2V       &   6708 & 4.32 & $-$0.38 \\
HD130945 & F7IV      &   6431 & 4.06 &    0.06 \\
HD132375 & F8V       &   6344 & 4.25 & $-$0.05 \\
HD134083 & F5V       &   6632 & 4.50 &    0.10 \\
HD136064 & F9IV      &   6140 & 4.02 & $-$0.05 \\
HD137052 & F5IV      &   6385 & 3.91 & $-$0.12 \\
HD139457 & F8V       &   5941 & 4.06 & $-$0.52 \\
HD142357 & F5II-III  &   6450 & 3.30 &    0.20 \\
HD142860 & F6IV      &   6280 & 4.10 & $-$0.18 \\
HD144206 & B9III     &  11833 & 3.67 &    0.01 \\
HD144284 & F8IV      &   6309 & 4.13 &    0.20 \\
HD145976 & F3V       &   6720 & 4.10 &    0.01 \\
HD150012 & F5IV      &   6380 & 3.80 &    0.05 \\
HD155646 & F5IV      &   6179 & 3.92 & $-$0.14 \\
HD157373 & F4V       &   6420 & 4.07 & $-$0.48 \\
HD157856 & F3V       &   6309 & 3.93 & $-$0.18 \\
HD159332 & F6V       &   6184 & 3.85 & $-$0.23 \\
HD161149 & F5II      &   6600 & 2.95 &    0.55 \\
\hline\hline
\end{tabular}
\end{center}
\end{table}

\subsection{The wavelength sequence, line ratios and Lick-like indices}

We explored all possible combinations of the indices defined by Rose
(1994), which are in the form of flux ratios at the central
wavelengths of absorption lines or at pseudo-continuum loci, and
the Lick/IDS indices (e.g., Worthey et al. 1994; Trager et al.
1998). We visually inspected the spectra of the template stars in
search of additional pairs of features that could display a trend
with gravity. At the end of the process, the indices that emerged as
best gravity diagnostics are (1) the combination 4289/4271 vs.
H$\gamma$/4325 from Rose's indices, for stars with $T_{\rm eff}\leq
7500$~K, and (2) the Lick index $H\gamma_{\rm A}$, for stars hotter than
8000~K.

\subsection{Analysis of line-depth ratios}

As an important preliminary step, we theoretically verified the
sensitivity of all of Rose's spectral features to surface gravity, in
particular those that he identified as discriminators of gravity.

In spite of the similar spectral resolutions, this verification
process is needed because Rose's spectra were not flux calibrated and,
therefore, subject to effects inherent to the particular instrumental
set-up he used. In other words, we have corroborated that the selected
indices indeed separate the effects of effective temperature from
those of gravity.

We calculated the line-depth ratios in a subsample of solar
chemical composition synthetic spectra of {\sc bluered}, after
properly degrading the grid to match the working resolution of
2.6~\AA. As mentioned previously, after exploring the full set of line
ratios, we identified the diagnostic diagram including 4289/4271
vs. H$\gamma$/4325 as the best fiducial combination for
differentiating amongst stellar luminosity classes.

\begin{figure}
\begin{center}
\includegraphics[width=4.0cm]{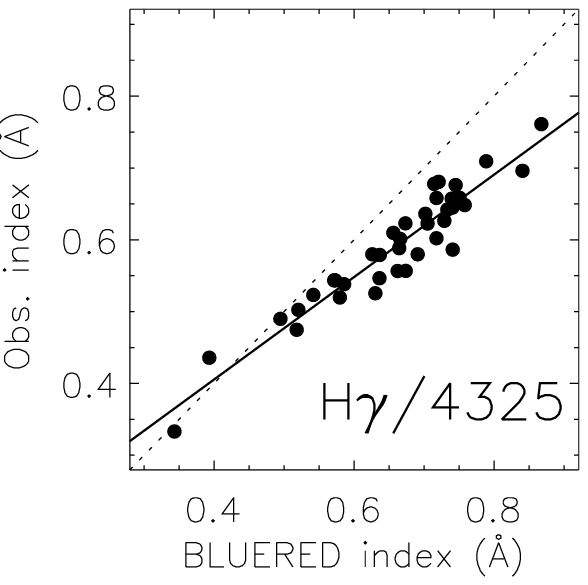}
\includegraphics[width=4.0cm]{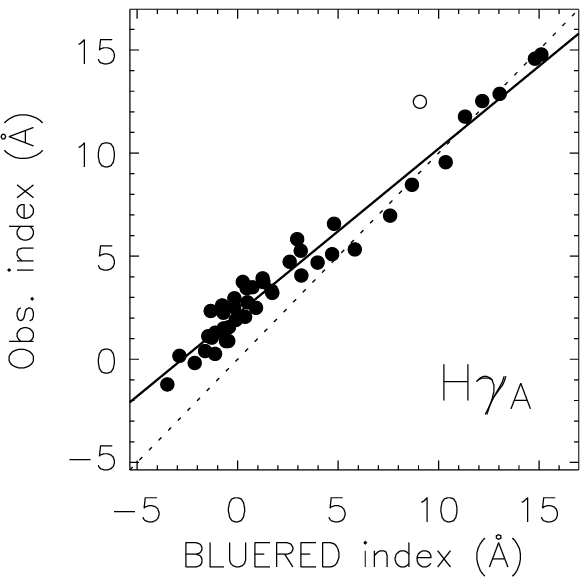}
\end{center}
\caption{Comparisons between the theoretical and empirical
indices for the sample of template stars. The empty circle in the
right-hand panel corresponds to an object deviating more than
3$\sigma$. This object was not taken into account for the
calibration. The dotted lines show the one-to-one correlations
whereas the solid lines denote the best fits.
\label{fig:tvse}}
\end{figure}

Once we have confirmed their sensitivity to gravity, the next
step is to transform the theoretical indices to our observational
system. For this calibration we compare the theoretical indices with
those measured from observed spectra. Note that theoretical values
were obtained from {\sc bluered} spectra after linearly
interpolating the set of parameters of the template stars
(Table~\ref{tab:refstars}). As an example of this comparison we show
in Fig.~\ref{fig:tvse} the correlations between the theoretical and
the empirical indices for H$\gamma$/4325 and H$\gamma_{A}$. This
figure indicates that a linear transformation of the form $index_{\rm
theor} = a + b \times index_{\rm obs}$, suffices to properly match the
theoretical and empirical indices.


The comparison resulted in the transformation coefficients
listed in Table~\ref{tab:coeff}, along with the root mean square
error.

\begin{table}
\begin{center}
\caption{Linear transformation parameters.\label{tab:coeff}}
\begin{tabular}{llll}
\hline\hline
Index & a & b & rms \\
\hline
4289/4271         & 0.385 & 0.605 & 0.017 \\
H$\gamma$/4325    & 0.120 & 0.713 & 0.026 \\
H$\gamma_{\rm A}$ & 2.21  & 0.80 & 0.74 \\
\hline\hline
\end{tabular}
\end{center}
\end{table}

\begin{figure}
\begin{center}
\includegraphics[width=8.5cm]{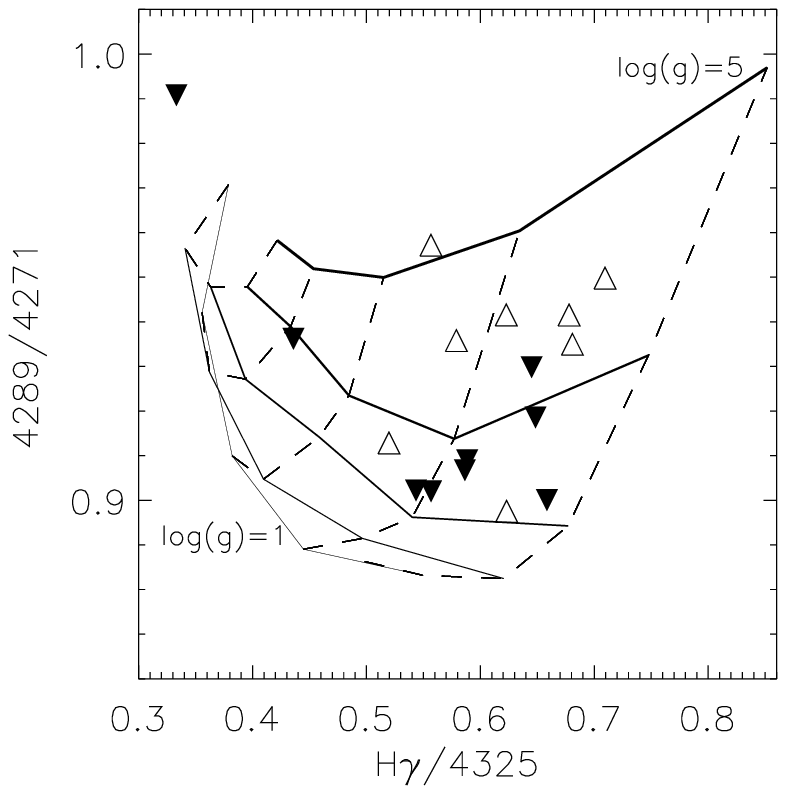}
\includegraphics[width=8.5cm]{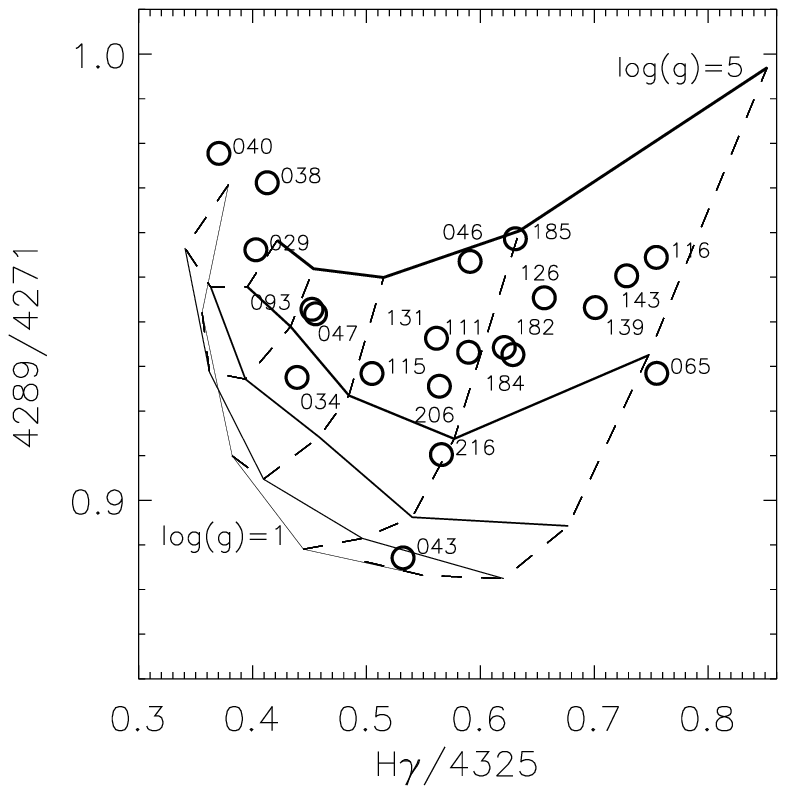}
\end{center}
\caption{Diagnostic diagram for Rose's indices 4289/4271 vs.
  H$_\gamma$/4325. Dashed lines indicate the iso-$T_{\rm eff}$ curves at
  6000, 6500, 7000, 7500, 8000, and 8500~K (from right to left), while
  the solid lines show iso-gravity trends at $\log{g}$=5, 4, 3, 2 and
  1~dex for the theoretically calculated indices. In the {\it top
  panel} the upward and downward triangles indicate, respectively,
  reference stars with gravities in the intervals
  $\log{g}$=4.0--4.5~dex and $\log{g}$=3.5--4.0~dex. In the {\it
  bottom panel} the positions of the BSs are marked with open circles,
  along with their identifications.
  \label{fig:rose}}
\end{figure}

In Fig.~\ref{fig:rose} we display the theoretical diagram (for
solar metallicity) for a set of different effective temperatures and
gravities, after application of the transformation described
above. The dashed and solid lines illustrate, respectively, the
iso-$T_{\rm eff}$ and iso-gravity curves. In the top panel we overplot
the loci of the template stars with $-0.15\leq$ [Fe/H] $\leq$+0.15
dex. In the bottom panel we show the same diagram and the positions
of the BSs.

From inspection of the panels in Fig.~\ref{fig:rose} we note the
following characteristics: up to an effective temperature of 7500~K,
the indices can clearly be used to separate among stars of the three
higher gravity bins, a sensitivity which appears enhanced for dwarfs
and subgiants. Most of the BSs show up in the interval
$\log{g}$=4.0--5.0. The only exceptions are BS034, BS043, BS065 and
BS216, for which their loci in the diagrams indicate gravities lower
than log$g$ = 4.0~dex. Interestingly, in the work of Mathys (1991) two
of these stars, BS034 and BS043, also turned out to be the
lowest-gravity objects, with $\log{g}$=3.79 and 3.44, respectively,
although the latter star might require a more detailed analysis since
it is part of a close pair (Girard et al. 1989).

Importantly, we note at this point that we have not (yet) attempted to
use these diagrams to derive values for the atmospheric parameters,
but instead we only provide an overall assessment of the gravity of
the objects. For a more quantitative evaluation, a detailed analysis
regarding the adequacy of the model spectra is necessary and still
beyond the scope of the present paper. At any rate, it is important to
mention that in spite of the potential problems associated with
theoretical spectra (see Bertone et al. 2008) the diagrams clearly
exhibit (even without prior knowledge of the atmospheric parameters)
that most stars with $T_{\rm eff} \leq 7500$~K have surface gravities
compatible with stars on the main sequence.

\subsection{The H$\gamma$ Lick-like index}

\begin{figure}
\begin{center}
\includegraphics[width=8.5cm]{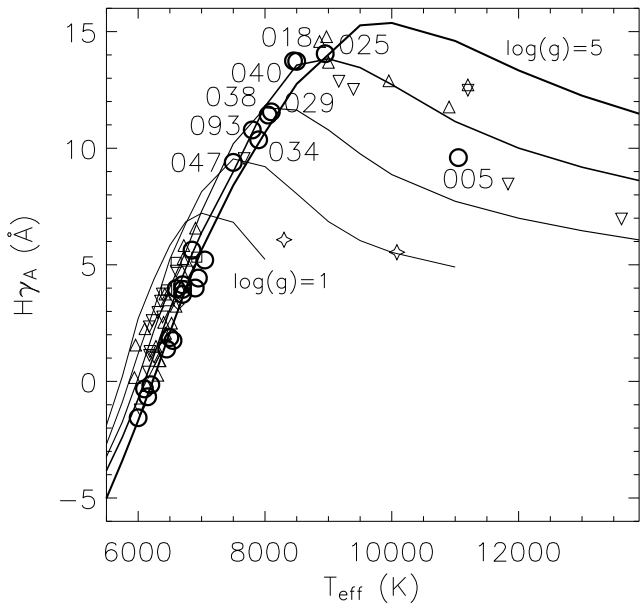}
\end{center}
\caption{Lick-like index H$\gamma_{\rm A}$ as a function of effective
  temperature. The solid lines show the theoretical iso-gravity curves
  from $\log{g}$=5 to 1~dex from solar metallicity spectra. Observed
  stars are marked with the same symbols as in Fig.~\ref{fig:rose}
  with the addition of squares marking objects with log
  $g$=2.5--3.5 dex and starred symbols denoting stars with log $g<$2.5
  dex.\label{fig:lick}}
\end{figure}

For the BS stars of higher temperatures we have measured absorption
indices of the hydrogen Balmer lines, as defined by Trager et
al. (1998). We have termed these indices 'Lick-like' since they have
not been transformed to the Lick system. The overall behaviour of the
indices associated with the H$\gamma$ and H$\delta$ lines in empirical
data has demonstrated that the indices barely depend on metallicity
and are very sensitive to gravity for stars with $T_{\rm eff} >
8500$~K. We have constructed a theoretical diagnostic diagram of
H$\gamma_{\rm A}$ vs. $T_{\rm eff}$ using {\sc bluered}. In a
similar fashion to the line-depth indices, we have calibrated
theoretical indices by comparing them to the indices measured in the
template stars. The results are included in Table~\ref{tab:coeff}. In
Fig.~\ref{fig:lick} we display the theoretical trends for solar
chemical composition, with the stars represented as in
Fig.~\ref{fig:rose}. In Fig.~\ref{fig:lick} we include the full
sample of template stars regardless of their chemical composition.
Note that the index values degenerate at low temperatures, whereas
stars are well separated at high temperatures, in particular BS005
(our hottest object). According to this diagram, the three hottest
stars have surface gravities in excess of $\log{g}$=4.0, although --
because of their temperature -- their loci in the diagram do not allow
us to precisely establish this parameter. The hot object BS005 appears
to have a gravity of about $\log{g}$=3.6, which is compatible with our
determination using the {\sc bluered} grid.

There are two objects, BS029 and BS038, that do not lie within the
physically expected regions in the two diagrams. For these stars,
Mathys (1991) determined effective temperatures consistent with our
determination, and gravities of $\log{g}$=3.91 and 4.14 for BS029 and
BS038, respectively.

Therefore, the spectral properties of BSs can be represented by main
sequence stars with the same photometric properties when modeling a
simple stellar population based on (photometric) observations of star
clusters.

\section{Comparison with previous work}

Previous studies of BSs in M67 include, for instance, Bruntt et
al. (2007) for BS018, BS025, BS034, BS038, BS040, BS047, and BS093,
based on asteroseismic analysis for $\delta$ Scuti pulsations, and
Shetrone \& Sandquist (2000) for BS043, BS046, BS139, and BS206, based
on abundance analysis. Mathys (1991) analysed 11 BSs in M67 and
studies on binarity are also available (see below). However, a
homogeneous survey of the full sample of BSs in M67 and complete
atmospheric parameter determinations have not yet been carried
out. Mathys (1991) presented a spectroscopic study of 11 BSs in M67,
and performed a detailed abundance analysis for F 153 and F 185. He
concluded that the effective temperatures and surface gravities of the
BSs in M67 were quite similar to those of normal main sequence stars
of the same spectral type.

There is an obvious difference between his method and ours. Based on
the photometric data from Mermilliod (1988), Mathys (1991) derived the
atmospheric parameters from the photometric measurements of the BSs in
the Str\"{o}mgren system, applying the relevant calibration to
determine the effective temperatures and surface gravities of B-, A-
and F-type stars using {\it{uvby$\beta$}} photometry (Moon \&
Dworetsky 1985).

\begin{figure}
\includegraphics[width=6.5cm]{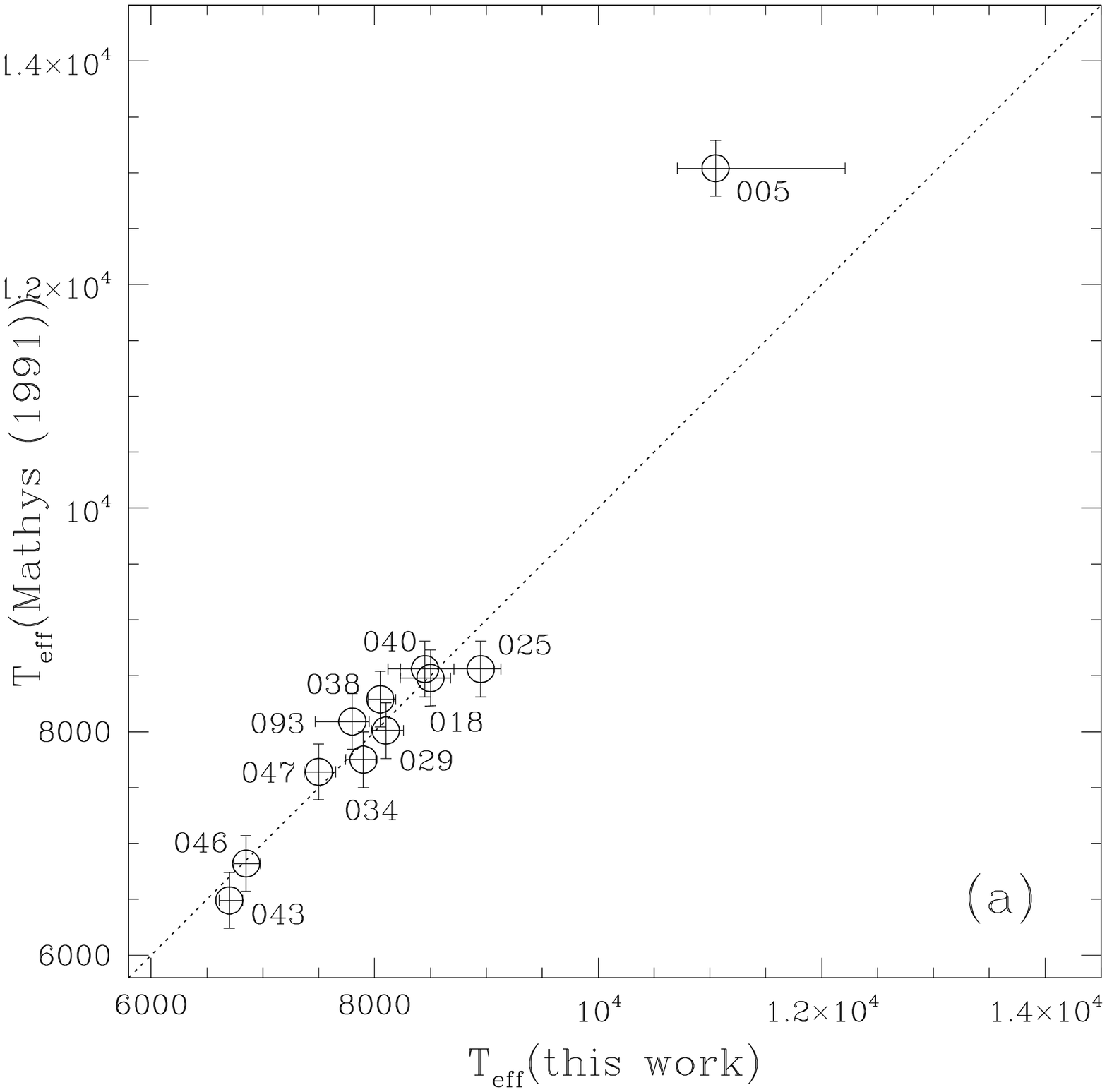}
\includegraphics[width=6.5cm]{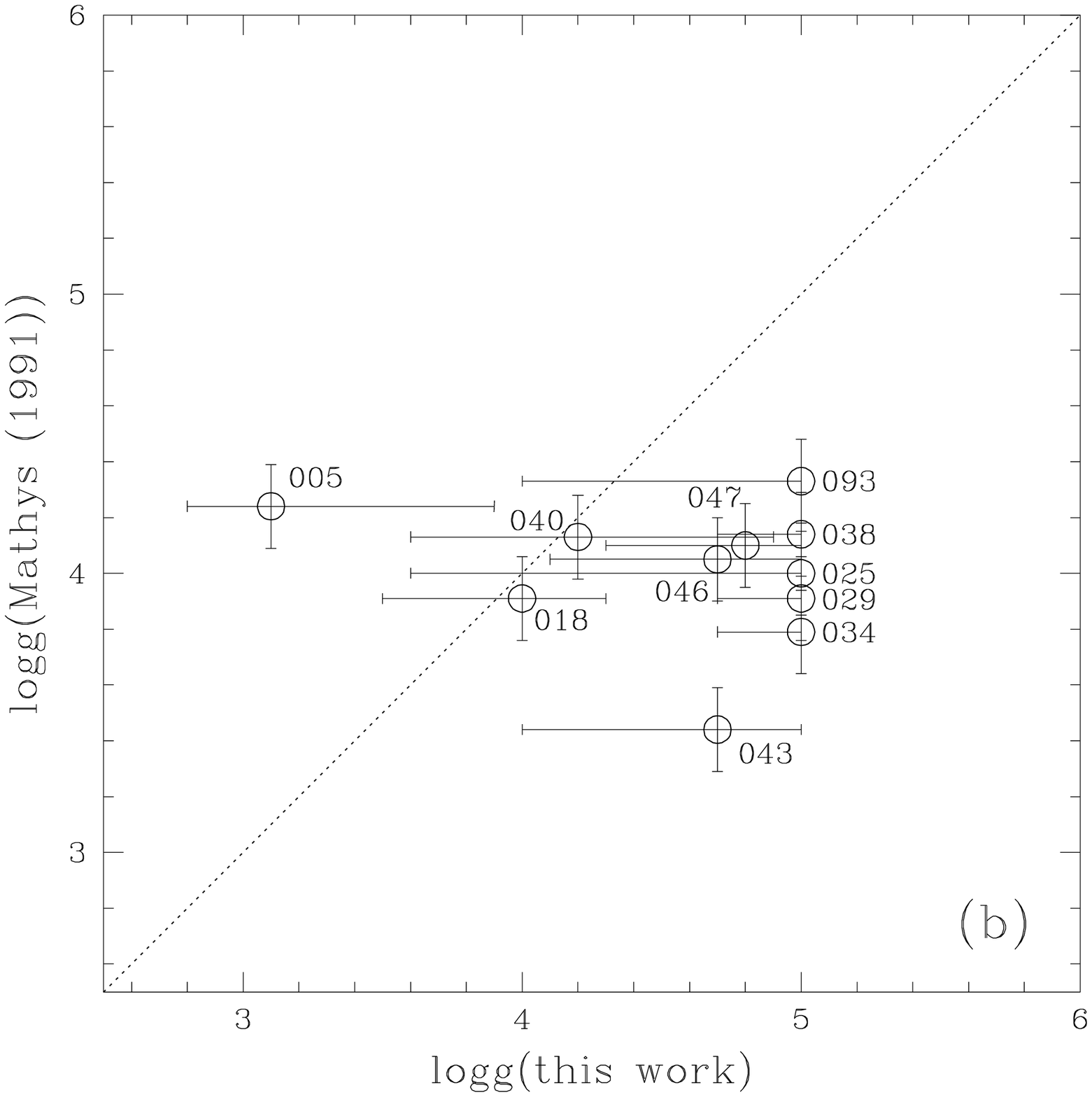}
\caption{Comparison between our results and those of Mathys
    (1991): (a) effective temperature; (b) surface gravity. The dotted
    lines are the one-to-one correlations. The open circles are the 11
    BSs (labeled by star IDs) in common between the present work and
    Mathys (1991).} \label{mathys}
\end{figure}

The effective temperatures derived from the present work (from
{\sc bluered}) for the 11 BSs in common are consistent with Mathys
(1991), as shown in Fig.~\ref{mathys}(a), and the surface-gravity
determinations are less conclusive for most of the BSs compared with
Mathys (1991), as shown in Fig.~\ref{mathys}(b). The error bars
(horizontal axes) in Fig.~\ref{mathys} were obtained based on {\sc
bluered} spectral fits, whereas the vertical error bars are from
Mathys (1991).

For the 11 BSs in common, the surface gravities of BS005, BS018,
BS025, BS038, BS040, BS046, BS047, and BS093 in both papers are in
fairly good agreement, considering the error bars. Large deviations
in surface gravities are found for the remaining three BSs in common
(BS029, BS034, BS043), as shown in Fig.~\ref{mathys}(b). Very likely,
the undoubted binary nature of these three objects is responsible for
the deviations between the two methods.

Indeed, there are eight objects in the list of BSs in M67 which are
likely binary candidates, based on previous observations. The BSs
identified as binaries are marked by asterisks in
Table~\ref{result}. The BS BS034 (S1284), for instance, is thought to
be a binary system in the final stages of mass transfer (Milone \&
Latham 1992; Zhang et al. 2005). Milone \& Latham (1992) considered
the dominant light contributor in BS034 to be the original primary
(now the secondary) with an orbital period of 4.18284 days and an
eccentricity of {\it e}=0.205. Based on high-precision radial velocity
measurements, they obtained a spectroscopic orbital solution for the
BS binary system. They supposed that the mass transfer began fairly
recently and that this BS was formed through stable mass transfer with
nearly 100 per cent efficiency. BS046 (S1082) was found to be a
complex unusual eclipsing binary system, or even a triple system of
which the SED could be explained by the sum of a close binary and
another main sequence star (van den Berg et al. 2001; Zhang et
al. 2005). BS029 (S1267), BS043 (S975), BS047 (S752), BS111 (S997),
and BS115 (S1195) were all identified as spectroscopic binaries with
long periods ranging from 800 to 5000 days (Latham \& Milone
1996). BS184 (S1036) was detected as a W UMa-type binary with a small
amplitude of light variations, and a strong but stable O'Connell
effect (Sandquist \& Shetrone 2003).

In our work, these binaries were easily fitted using model spectra of
single stars.  Although we cannot corroborate their binary nature, to
within the limited resolution of the observations, we can nevertheless
provide constraints on the BSs' spectral properties.

\section{Summary and Discussion}

This study represents the first attempt to derive the parameters of
the full sample of BSs in the old Galactic open cluster M67 (NGC2682)
in a homogeneous way. Low-resolution spectra of the sample of 24 BSs
in M67 were collected using the 2.12 m telescope of the Guillermo Haro
Observatory (Mexico). The entire data set was re-calibrated using the
BATC intermediate-band photometric system, in addition to the usual
relative calibration using standard stars, and was subsequently 
used for a comparison with three different stellar databases aimed at
studying their spectral properties in a systematic way. We found that
all objects have gravity values in agreement with the expected values
for objects in the hydrogen-burning stage.

Considering the original goal of our work, we conclude that, in
terms of spectroscopic properties at low resolution, the BSs can
indeed be represented by empirical or theoretical data of (or
compatible with) main sequence stars,
at least in a low density environment as in M67.

As a natural
extension to this, it is further concluded that when building up the
empirical SEDs of SSPs based on stellar clusters, the contributions
due to BSs can be accounted for using photometry and stellar spectral
libraries. This conclusion holds at least at low and intermediate
spectral resolution.

Limited by the spectral resolution of the current observational data
set, it is not possible to assess binarity and the formation mechanism
of the sample of BSs in M67. We anticipate that a detailed chemical
abundance analysis at high resolution will show signatures of these
dynamical and physical processes. Therefore, the current work serves
as a valuable starting point.

\section*{acknowledgments}
We thank the anonymous referee for rapid and useful comments. We
would like to thank the National Science Foundation of China (NSFC)
for support through grants 10573022, and the Ministry of Science and
Technology of China through grant 2007CB815406. MC and EB would like
to thank CONACYT through grants SEP-2005-49231 and SEP-2004-47904.
We would like to thank Richard de Grijs for language proof reading the paper.

\end{document}